\documentclass[twocolumn,fleqn,natbib]{svjour2}    % twocolumn

\usepackage{graphicx}
\usepackage{mathptmx}      % use Times fonts if available on your TeX system

\journalname{Experiments in Fluids}

\begin{document}

\title{Near-wall velocity measurements by Particle-Shadow-Tracking}

%\titlerunning{Short form of title}        % if too long for running head

\author{Pierre Lancien         \and
        \'{E}ric Lajeunesse        \and
        Fran\c{c}ois M\'{e}tivier      \and
}

%\authorrunning{Short form of author list} % if too long for running head

\institute{P. Lancien \and E. Lajeunesse \and F. Metivier \at
              Laboratoire de Dynamique des Fluides G\'eologiques\\ 
              Institut de Physique du Globe de Paris \\
              4, Place Jussieu\\
              75252 Paris cedex 05, France               
}

%\date{Received: date / Accepted: date}
% The correct dates will be entered by the editor

\maketitle
\begin{abstract}
We report a new method to measure the velocity of a fluid  in the vicinity of a wall.  The method, that we call Particle-Shadow Tracking (PST), simply consists in seeding the fluid  with a small number of fine tracer particles of density close to that of the fluid.   The position of each  particle and of its shadow on the wall are then tracked simultaneously, allowing one to accurately determine the distance separating tracers from the wall and therefore to extract the  velocity field.  We present an application of the  method to the determination of the velocity profile inside a laminar density current flowing along an inclined plane.  
\end{abstract}

\section{Introduction}
\label{intro}
Measuring the velocity of a fluid in the vicinity of a wall is relevant to a great number of fundamental and applied investigations such as understanding the structure and dynamics  of  boundary layers \cite{Townsend-1976,Alfredsson-1988,Buschmann-2003}.  In practice, such measurements are difficult as they require to achieve a high accuracy on both the velocity and the distance to the wall, the proximity of which makes unfortunately difficult to reach. Optical  methods  such as Particle Image Velocimetry (PIV) or  laser anemometry usually fail to meet these two criteria when operating too close to a wall and are therefore inappropriate for such measurements \cite{Adrian-1991,Somandepalli-2004}. 

In this paper, we present a new and simple experimental technique developed to perform local measurement of the velocity field of a fluid near a wall. This method, that we call Particle-Shadow Tracking (PST), consists in seeding the fluid with a very low number of fine tracer particles of density close to that of the fluid.  Simultaneous tracking of each particle and its shadow on the wall allows us to accurately determine the distance separating tracers from the wall and therefore to extract the velocity field.

The paper is organized as follows. The technique is presented and discussed in section \ref{sec:1}. Section 3 describes an application of the method to the measurement of the velocity profile of a laminar density current flowing along an incline plane. We then make some brief suggestions about how the technique may be further developed and conclude.

\section{The PST method}
\label{sec:1}
	
\begin{figure*}
\includegraphics[width=0.70\textwidth]{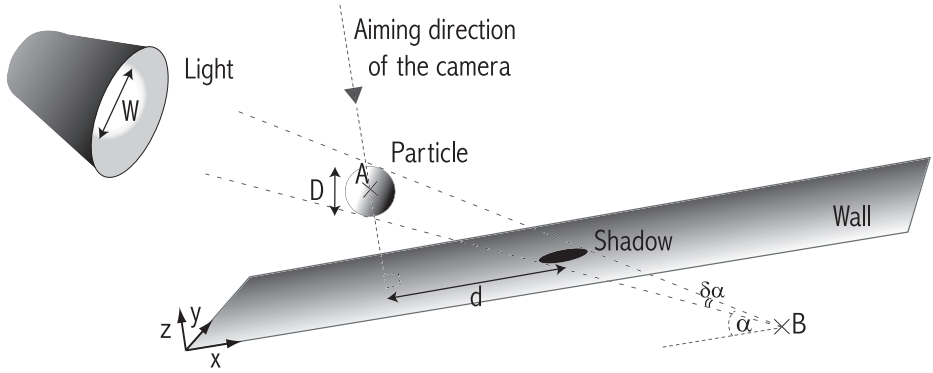}
\caption{Schematic of the PST method}
\label{fig:schemapst}
\end{figure*}

For the sake of simplicity, we will illustrate the PST method in the case of a fluid flowing  above a inclined wall as sche\-med on figure \ref{fig:schemapst}. The  $x$ and  $y$ axes are  oriented respectively along and transverse to the flow direction. The $z$ axis is normal to the wall. The fluid of density $\rho$ and viscosity $\eta$  is seeded with a small number of fine tracer particles of diameter $D$. Ideally the best is to use neutrally buoyant particles, that is of density $\rho_p=\rho$. In practice, a perfect density matching is difficult to achieve. But, as discussed hereafter, the use of tracer particles of density  slightly larger than that of the fluid can be  a plus for the method.

A punctual light source enlightens the wall under a small incidence angle $\alpha$.  A camera oriented perpendicularly to the plane is used to acquire digitized images of the wall at regular time intervals.  The images are then processed to extract the lateral coordinates ($x_p$,$y_p$) of each particle and ($x_s$, $y_s$)  of the shadow it projects on the wall (see fig. \ref{fig:schemapst}). The third coordinate $z_p$ of the particle is then deduced from the distance $d$ separating the particle from its shadow:
\begin{equation}
\label{zp}
z_p = \tan{\alpha} \; d \;  \approx \alpha \, d
\end{equation}

Measurement of $x_p(t)$, $y_p(t)$ and $z_p(t)$   at regular  time intervals therefore allows us to determine the velocity field along the particle trajectory: 
\begin{equation}
\label{u}
u(x_p,y_p,z_p) =  \frac{x_p(t+\delta t) - x_p(t)}{\delta t}   
\end{equation}
where $\delta t$ is the time step separating the acquisition of  two successive frames. The complete velocity profile $u(x,y,z,t)$ is extracted by repeating these measurements on a large number of particles whose trajectories cover the whole area of interest.

In terms of accuracy, it follows from equations (\ref{zp}) and (\ref{u}) that:
\begin{eqnarray}
\label{precis}
\Delta z_p & = & \alpha R \\  
\Delta u & = & 2  \frac{R}{\delta t}   
\end{eqnarray}
where $R$ is the  spatial resolution of the image (i.e. the  pixel size).

As for all velocimetry method based on seed particles, an important issue is to ensure that particles move at the same velocity than the fluid. In the limit of small particle Reynolds number, the difference between fluid and particle velocity is given by \cite{Adrian-1991}:
\begin{eqnarray}
	\left|v-u\right| = \frac{\rho_{p} {D}^2 \left|\dot{v} \right|}{36  \eta}
\end{eqnarray}
where $v$ is the velocity of the particle, $u$ that of the fluid and $\dot{v}$ is the typical particle acceleration. If the flow is steady and quasi-parallel, it follows from the above equation that $v$ is very close to $u$.  For a non steady flow, the effect of slip needs to be evaluated from the above equation in order to estimate the accuracy of the velocity measurement.

In practice, Particle-Shadow tracking forces velocity measurements to remain concentrated below a maximum distance to the wall for two  reasons. First of all, both the particle and its shadow have to be present simultaneously inside the zone imaged by the camera. It follows from equation (\ref{zp}) that the largest wall-particle distance that can be detected is of the order of:
\begin{equation}
\label{eq:zmax1}
z_1 \approx S.\alpha
\end{equation}
where $S$ is the typical length of the zone imaged by the camera. 
  
The second limiting factor  is illustrated on figure~\ref{fig:schemapst}.  In practice, a light source is not punctual. As a result,  the sha\-dow cone generated by a particle A extends up to a point B,  beyond which only  penumbra remains. If B is located above the wall, the shadow poorly contrasts on the wall so that it cannot be detected. Distance AB can be estimated by:
\begin{equation}
\label{eq:distAB}
\| AB \| \approx \frac{D}{\delta\alpha}
\end{equation}
where $D$ is the particle diameter. $\delta\alpha \approx W/L$ is the source angular diameter, where $W$ is the light source diameter and $L$ is the distance separating the light source from the measurement point. From figure~\ref{fig:schemapst} it follows that the shadow is not detectable anymore as soon as the distance of the particle to the wall is larger than:
\begin{equation}
\label{eq:zmax2}
z_{2} \approx D\frac{\alpha}{\delta\alpha} = \frac{D \alpha L}{W}
\end{equation}
Although receding the light source reduces $\delta\alpha$  and increases $z_{2}$, the counterpart is   a loss of luminosity degrading the signal to noise ratio during treatment.  Another alternative is to use a collimated light source instead of a punctual one. In that latter case, $z_{2}$ depends on  the parallelism uncertainties $\delta \alpha'$ of the collimated light: $z_{2} \approx D \alpha / \delta\alpha'$  \cite{Adrian-1991}.  

A key point of the method is to determine the shadow associated to a given particle. This can become difficult when the number of particles visible on a frame is too large. A simple way to address this problem is to use tracer particles of density  slightly larger than that of the fluid so that they slowly settle. When a  settling particle reaches the wall, it coincides with its shadow allowing us to identify it without any ambiguity. The particle and its shadow are then tracked by playing the movie backwards. This however  introduces a third  maximum distance to the wall  as the particle must remain in the field of the camera until it reaches the wall. The time necessary for the particle to settle from a height $h$ is:
\begin{equation}
\label{eq:tsettling}
t = \frac{h}{V_s} 
\end{equation}
where the settling velocity $V_s=(\rho_p -\rho) .g.D^2 / 18 \eta$ is estimated from the Stokes velocity (Lamb 1945). The horizontal distance covered by the tracers during this time is of order: 
\begin{equation}
\label{eq:distsettling}
\phi = \frac{h U}{V_s}
\end{equation}
where $U$ is the typical flow velocity.  The particle must remain in the field of the camera until its settling, which imposes $\phi<S$ and determines a third boundary for the profile height of the order of:
\begin{equation}
\label{eq:zmax3}
z_{3} \approx \frac{S V_s}{U}
\end{equation}

From the above discussion, it follows that the maximum wall-distance $z_{max}$ which can be explored by PST is given by the minimum between $z_{1}$, $z_{2}$ and possibly $z_{3}$.   For $z > z_{max}$, PST fails and the velocity profile has to be measured using classical techniques such as PIV.

\section{Application}
\label{sec:2}

\begin{figure}
\includegraphics[width=0.45\textwidth]{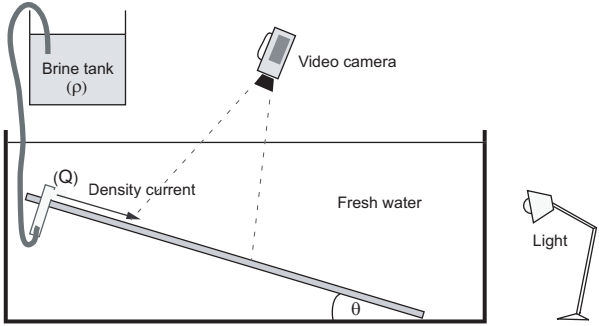}
\caption{Setup for the density current experiment}
\label{fig:setup}
\end{figure}

We first developed the PST method in order to measure the velocity profile of a dense current running out on a slope \cite{Lancien-2005a,Lancien-2005b}.  We were particularly interested in the profile in the immediate vicinity of the wall as our goal was to measure the shear stress applied on the latter by the current. We will therefore illustrate  PST for this particular application. The corresponding experimental setup is sketched on figure \ref{fig:setup}. It consists of a $100 cm \times 50 cm$ incline, immersed in a $200 cm \times 50 cm \times 50 cm$ flume filled with fresh water. The flow is generated by injecting a brine of density $\rho_b$ larger than that of fresh water at a constant flow rate  $Q$ from the top of the ramp. The resulting gravity current is laminar, steady and varies slowly along the $x$ axis \cite{Lancien-2005a,Lancien-2005b}.

Our camera definition is $700 \times 570$ pixels, the width of the imaged zone is $S = 7 cm$ and the spatial resolution of the camera is  $R = 100 \mu m$/pixel. The acquisition rate is 25 frames $s^{-1}$ corresponding to $\delta t = 40 ms$.  The experimental plane is enlightened with a projector under an incidence angle $\alpha = 20^{\circ}$.  The angular diameter of the lamp is $\delta \alpha \approx 0.4^{\circ}$.  The tracers are fine plastic particles (Rilsan) with a characteristic size of $30 \mu m$.  Their density  $\rho_p =1080 Kg.m^{-3}$ is slightly larger than that of the brine. As a result,  they slowly settle so that one particle allows us to explore a  wide range of $z$.  Note that,  as our  flow is steady and varies slowly along the $x$ axis,  tracking only one particle is enough to establish the velocity profile $u(z)$.

\begin{figure}
\includegraphics[width=0.45\textwidth]{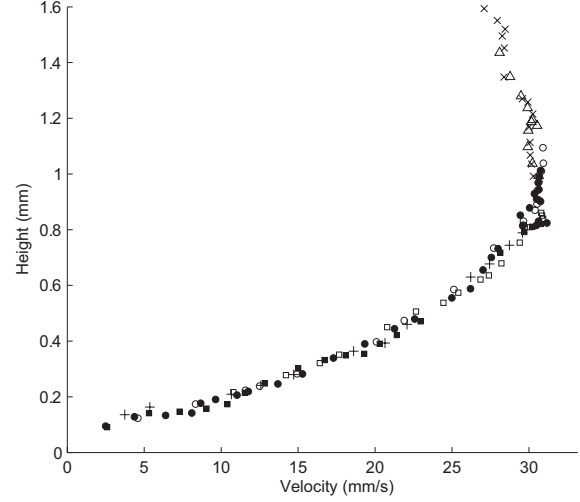}
% figure caption is below the figure
\caption{Near-wall velocity profile of the density current, measured $0.3 m$ downslope of the entrance with PST. Each symbol correspond to a different particle. $\rho = 1025 Kg.m^{-3}$,$ Q = 0.95 mL.s^{-1}$, $\theta = 16^{\circ}$.}
\label{fig:pro}
\end{figure}

Equations (\ref{eq:zmax1}), (\ref{eq:zmax2}) and (\ref{eq:zmax3}) lead to: $z_{1} \approx 10 mm$, $z_{2} \approx 1.5 mm$ and $z_{3} \approx 5 mm$.  In this experimental configuration, the maximum wall-distance $z_{max}$ which can be explored by PST is set by the size of the light source which forbids to measure velocity  above \- $z_{max} = z_{2} \approx 1.5 mm$.   Figure \ref{fig:pro} shows typical velocity measurements obtained by applying PST to 5 different particles located in the same area.  All data collapse on the same velocity profile $u(z)$. The  flow that we consider in this section is  laminar and stationary. By performing several repeated measurements  and  performing slide averages of the data, we were able to establish the velocity profile with an accuracy of $0.3 mm.s^{-1}$ and a vertical resolution of $30 \mu m$ (that is inferior to the pixel size) up to a maximum distance  $z_{max} \approx 1.5 mm$.

\begin{figure}
\includegraphics[width=0.45\textwidth]{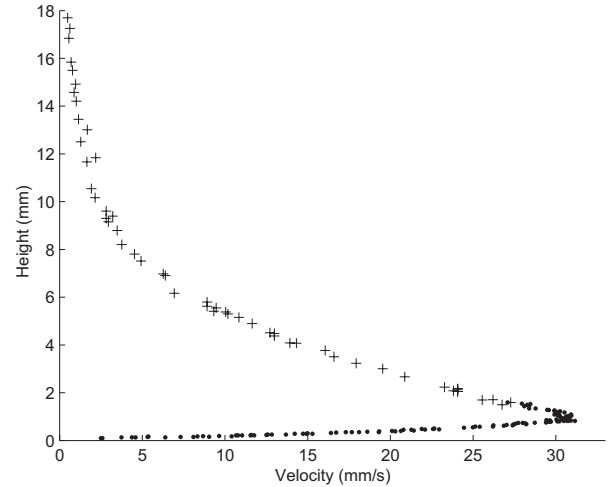}
% figure caption is below the figure
\caption{Complete velocity profile. PST data from figure \ref{fig:pro} (dots) are completed using classical side-view particle tracking (crosses).}
\label{fig:profil}
\end{figure}

Above $z \approx 1.5mm$, the velocity profile was measured with a more classical particle tracking  technique using side-views of the flow acquired by a camera placed on the side of the tank.  The lower and upper part of the resulting velocity profile are perfectly consistent and overlap in the vicinity of $z \approx 1.5 mm$,   as visible on figure \ref{fig:profil}. To our knowledge, this is the first measurement of a complete  velocity profile obtained for laminar density current experiments.
 
\section{Conclusion}
\label{sec:4}

We have described  a new method to measure near-wall velocity profiles. This so-called Particle-Shadow Tracking me\-thod (PST) consists in seeding the fluid with a very low number of fine tracer particles. Tracking simultaneously  both the particles and their shadow permits to measure the velocity profile in the direction transverse to the wall.  As illustrated through an example, this new method is cheap,  simple and  accurate.  Measurements are however restricted  in a layer at the wall of thickness  which mainly depends on light conditions, flow velocity and particles size and density.  Determination of the velocity profile out of this layer have to be performed using more classical techniques such as PIV.  PST should be therefore considered as a complementary method particularly adapted to the investigation of boundary layers.

Using two light sources with two different incidence angles, a small one and a large one, might improve PST method by extending the thickness of the measurement layer. In fact, each particle would therefore project two different shadows on the wall. When possible, the particle-wall distance  would be calculated from the most distant shadow. When the latter is not in the field of view, \- particle-wall distance would be calculated from the nearest shadow.   It would thus be  possible to increase  the thickness of the measurement layer without degrading the vertical resolution.

Our motivation to develop the PST method was indeed to measure the velocity profile inside a laminar density current. Therefore we did not apply the method to a turbulent flow. In principles, PST (that is  simultaneous tracking  of the position of a particle and its shadow) should work in a turbulent flow.  Of course, in practice, applying the PST technique to a turbulent flow would be more difficult: a higher frame rate is needed and the size and density of the particles need to be adapted to prevent slip effects. 

%\begin{acknowledgements}
%If you'd like to thank anyone, place your comments here
%and remove the percent signs.
%\end{acknowledgements}

% BibTeX users please use one of
%\bibliographystyle{spbasic}      % basic style, author-year citations
%\bibliographystyle{spmpsci}      % mathematics and physical sciences
%\bibliographystyle{spphys}       % APS-like style for physics

\end{document}